\documentclass[doublecol]{epl2} 
\usepackage{amsmath}
\usepackage{amssymb}
\usepackage{float}
\usepackage{color}
\usepackage{dcolumn}
\usepackage{bm}
\usepackage{xcolor}
\usepackage{multirow}
\usepackage{adjustbox}
\usepackage{longtable}
\usepackage{comment}
\usepackage{array}
\usepackage[colorlinks,citecolor=blue,linkcolor=red,anchorcolor=blue,filecolor=blue,urlcolor=blue]{hyperref}

\title{Quest for a Universal Cluster Preformation Formula: \\ A new paradigm for estimating the cluster formation energy}
\shorttitle{Title} 

\author{Joshua T. Majekodunmi\inst{1} \and Raj Kumar\inst{2} \and M. Bhuyan\inst{3}}
\shortauthor{Majekodunmi, Kumar and Bhuyan}

\institute{                    
\inst{1} Institute of Engineering Mathematics,  Universiti Malaysia Perlis, Arau, 02600, Perlis, Malaysia\\
\inst{2} School of Physics and Materials Science, Thapar Institute of Engineering and Technology, Patiala, Punjab 147004, India\\
\inst{3} Center for Theoretical and Computational Physics, Department of Physics, Faculty of Science, University of Malaya, Kuala Lumpur 50603, Malaysia
}

\abstract{
This study presents a holistic picture of the preformation of nuclear clusters with credence to the kinematics of their emissions. Besides the fitting of the preformation formula to reproduce the experimental half-lives, we have investigated the interrelationship between the parameters involved in the cluster decay process for medium, heavy and superheavy nuclei. Based on the established conceptual findings, we propose a new cluster preformation probability ($P_0$) formula that incorporates all influential parameters of the cluster radioactivity and thus has an edge over the existing formulae in the literature. Further, we hypothesize that a fraction of the decay energy is needed for cluster formation within the parent nucleus. The proposed formula opens a new paradigm to separately estimate the energy contributed during the cluster formation from its emission and thus shows that the contribution of the Q-value splits into three major parts accounting for the energy contributed during the cluster preformation, its emission and recoil of the daughter nucleus. Moreover, the expression $P_0$ is adept at accommodating the theorized concept of heavy particle radioactivity (HPR). The result reveals that, like $\alpha$-decay, a proper estimation of the $P_0$ and $Q$-value in the cluster studies are enriched with qualitative information about the nuclear structure. However, from the analysis, the Geiger-Nuttall law is not the best compromise in the clustering due to the non-linearity between $\log_{10}T_{1/2}$ and $\sqrt{Q}$, unlike in $\alpha$-decay. We have demonstrated that with the inclusion of the proposed formula, the half-life predictions from both microscopic R3Y and phenomenological M3Y NN potentials closely agree with the available experimental data and that the slight variation can be traced to their peculiar barrier characteristics.}

\begin{document}

\maketitle
\section{Introduction}
Cluster radioactivity is a spontaneous disintegration/decay of radioactive nuclei where certain clusters heavier than $^{4}$He are emitted. Its discovery began with the theoretical prediction of Sandulescu \textit{et al.} \cite{sand80}. Subsequently, Rose and Jones \cite{rose84} detected it in an experiment in which $^{14}$C was emitted from $^{223}$Ra. Thereafter, the emission of heavier clusters ranging from  $^{20}$O - $^{34}$Si has been observed from several parent nuclei. Usually, such decays in the trans-lead region lead to the formation of daughters at $^{208}$Pb shell closure and its neighbours \cite{bone01,bone07,gugl08}. Sequel to the experimental detection, a few research works have been tuned towards understanding the theoretical underpinnings of the cluster emissions. From the theoretical front, the models employed in cluster emissions can be grouped into two, viz. the fission$-$ and $\alpha$- decay-like models, based on their treatment. In fission models, clusters are assumed to be formed during the separation or deformation process of the decaying parent as it penetrates the nuclear-Coulomb potential barrier. Examples of such models include the analytic super asymmetric fission model (ASAFM) of Poenaru \textit{et al.} \cite{poen85,poen86,poen12,poen10}. On the other hand, $\alpha$- decay-like models consider the pre-existence of clusters within the parent nuclei, before emission e.g. the quantum mechanical fragmentation theory \cite{maruhn74,fink74,guptaprl75} based performed cluster-decay model (PCM) of Gupta \textit{et al.}  \cite{gupt94,sing11,jos22a}. Thus, it incorporates the spectroscopic or preformation factor $P_0$ that assimilates the structural properties of the decaying parent nucleus. Hence, the estimation of $P_0$ is the most crucial problem in $\alpha$ decay-like models, which recognize a cluster as an entity formed within the nucleus preceding its emission and also depend on the structure of the decaying nuclei \cite{zhan08}.

From the microscopic perspective, it is elusive to obtain the exact value of $P_0$  due to the complexities associated with the nuclear many-body problem. Although the value of $P_0$ in the fission models is generally taken as 1 (unity) to predict the experimental half-lives, small discrepancies are usually inevitable which points toward a basic conceptual omission. These discrepancies are good determinants of the preformation probability \cite{zhan08,alsa16}. $P_0$ could be several orders of magnitude below unity, especially in cluster emissions \cite{mali89,phoo17}. Besides, the use of a constant preformation probability for all the nuclei in a specified region as those of the cluster model in Refs. \cite{buck89,buck92,xu06} could somewhat compromise the microscopic information about their respective nuclear structure. As such, Blendowske and Walliser introduced a phenomenological mass-dependent formula for the cluster preformation \cite{blen88}, which substantiates the experimentally observed cluster emission and asserts that $P_0$ can be estimated with realistic values. However, the limitation in Blendowske and Walliser's prescription is that its applicability is constrained to the mass of the cluster i.e. $A_c\leq 28$. For example, the formula becomes ineffective when considering some established concepts like spontaneous fission and heavy-particle radioactivity (HPR), where $Z_p=Z_c + Z_d >110$ with daughters $^{208}$Pb and those in its vicinity \cite{poen11,sant17,srid20,naga21}. In addition to these, the formula considers the mass of clusters only, thereby disregarding other nuclear properties that are subjected to its formation and emission. Later on, Wei and Zhang proposed a series of expressions \cite{wei17} showing the individual relationship between the cluster preformation and the mass, charge and neutron number of the parent, cluster and daughter nuclei with a reasonable fit to the experimental data. However, it relegates the dependence of the clusters on mass and charge asymmetry as well as the decay energy for their emissions which are known to influence the cluster decays \cite{bala04}. Moreover, the expressions show no clear correlations or interrelationships among the participating quantities. The Q-value-based formula of Santhosh $\&$ Tinu \cite{sant21c} appears to be more logical since it is predicated on the energy released during the decay process. It systematically incorporates the contributions of the parent, cluster and daughter nuclei. From the Chi-square ($\chi^2$) test, its deviation is estimated as 0.3293. Nonetheless, it is not a microscopic theory and its application is restricted to the heavy and superheavy mass region of the nuclear chart. The major drawback of the Q-value-based formula is that it gives no distinction between two (or more) systems with the same Q-value and hence, it predicts the same $P_0$ value regardless of their respective peculiarities. A vivid example of two systems with the same Q-value is given in  Ref \cite{yahy22} where $^{247}_{100}$Fm (at angular momentum $\ell=0$) and $^{241}_{99}$Es ($\ell=1.0$) has a Q-value of 8.26 MeV.

One could contemplate a better $P_0$ formula that harnesses the iso-spin dependence as those in Ref.\cite{sant21c} with $\chi^2$ deviation of 0.1561. This makes allowance for the situation where different clusters are being emitted from the same parent nucleus and it is possible to inspect the structural information of the emitted clusters, as well as their respective daughter nucleus \cite{bhu11,jos22b}. On the other hand, the decay energy is omitted here, which compromises the physics of the cluster emission process, since exothermicity, $Q>0$ cannot be undermined in any decay process. A few other $P_0$ formulae exist in literature \cite{sant21c} that fall under the above-discussed flaws. Particularly, the equations in Refs. \cite{sant21c} is based on the modified generalized liquid drop model (MGLDM) with elliptical lemniscatoid geometry. However, it has been lately shown by Poenaru \textit{et al.} that an important drawback of this geometry is the ``existence of a cusp at the intersection plane of the two fragments" which informs a back motion of their relative distance \cite{poen17}. The main drawback of all these formulae is that much emphasis and effort was given to fitting their respective expressions with some arbitrary values rather than investigating the characteristics and mechanism involved in the cluster preformation and its emission. It is pertinent to note that obtaining a reasonable fit is somewhat trivial as compared to understanding the physics of the underlying mechanism of a concept like a cluster preformation. These constitute our quest and motivation for the present study.

Here, we have employed the performed cluster-decay model (PCM), where the preformed cluster undergoes a quantum tunnelling process across the potential barrier formed by the interplay of the nuclear-Coulomb potentials \cite{gupt88,jos22a}. The nuclear potential is generally estimated by phenomenological \cite{quen78,horn75} or microscopic approaches \cite{schu16,vaut72}, which is the most important ingredient in cluster radioactivity studies. In this study, the recently developed R3Y nucleon-nucleon (NN) potential \cite{sing12,jos22b} which is derived from the relativistic mean-field (RMF) Lagrangian using the NL$3^*$ parameter set \cite{lala09} is used for the analysis. The results are also compared with its cognate phenomenological M3Y NN potential \cite{satc79}. The RMF approach is well-known for its considerably huge success in describing the ground and intrinsic excited-state properties of nuclei. Details and well documentation on the relativistic energy density functional for a variety of parameterizations and their limits can be found in Ref. \cite{dutra14} and the references therein. Besides the $P_0$, another input for the PCM is the barrier penetration probability $P$ which is calculated by employing the WKB approximation. The neck-length parameter of the PCM is fixed at $\Delta$R = 0.5 fm for M3Y and $\Delta$R = 1.0 fm for R3Y which is suitable for cluster decays in terms of their respective barrier properties \cite{jos22b,kuma12c,kuma12s}. In this work, the decay half-lives of $^{14}$C, $^{20}$O, $^{23}$F, $^{22}$Ne, $^{24}$Ne,$^{26}$Ne, $^{28}$Ne, $^{30}$Mg and $^{34}$Si clusters decaying from various heavy nuclei with the mass of the parent nuclei lying within the range $221\leq A\leq242$ are considered.

\section{Theoretical formalism and Discussions}
From the relativistic Lagrangian \cite{ring96,sero92,rein89,lala09}, one can obtain the fields for each meson using the Klein-Gordon equations \cite{ring96} and further considering the limit of a single-meson exchange for a static baryonic medium within the relativistic mean-field approach \cite{ring96,sero92,rein89,lala09}, the contributions of scalar and vector fields are used to derive the R3Y NN potential \cite{sing12,jos22b}. Similarly, the M3Y NN potential stems from the fitting of G-matrix elements predicated on Reid-Elliott soft-core NN-interaction \cite{satc79} in an oscillator basis. The R3Y and M3Y NN potential together with the RMF densities for the cluster(c) and daughter nuclei (d) undergoes the double folding procedure \cite{satc79} which yields the nucleus-nucleus potential. The nuclear potential $V_n(R)$ combines with the Coulomb potential $V_C(R) =\frac{Z_{c}Z_d}{R}e^2$ and results in the total interaction potential $V_T(R)$, which is used in the calculation of the WKB  barrier penetrability. It is worth mentioning that the three-step barrier penetration process \cite{jos22b} in the PCM framework is adopted here. The decay constant $\lambda$ and the half-life $T_{1/2}$ in the PCM take the form \cite{mali89},
\begin{equation}
     \lambda= \nu_{0} P_0 P,\hspace{0.5cm} T_{1/2}=\frac{\ln2}{\lambda}. \label{halfl}
\end{equation}
The $P_0$ and $P$ are the preformation probability and penetration, respectively. The assault frequency $\nu_0$ is calculated as 
 \begin{equation}
    \nu_{0}=\frac{\mbox{ velocity }}{2R_0}=\frac{\sqrt{2E_{c}/\mu}}{2R_0}, \label{afreq}
\end{equation}
where $R_0$ refers to the radius of the parent nucleus and $E_{c}$ is the kinetic energy of the emitted cluster. The Q-values are estimated using the  ground state binding energies \cite{wang17} from the expression
 \begin{equation}
     Q=BE_d+BE_c-BE_p \label{qval},
 \end{equation}
where $BE_p$, $BE_d$ and $BE_c$ are the binding energies of the parent, daughter nuclei and the emitted cluster, respectively.
\begin{figure}
\includegraphics[scale=0.325]{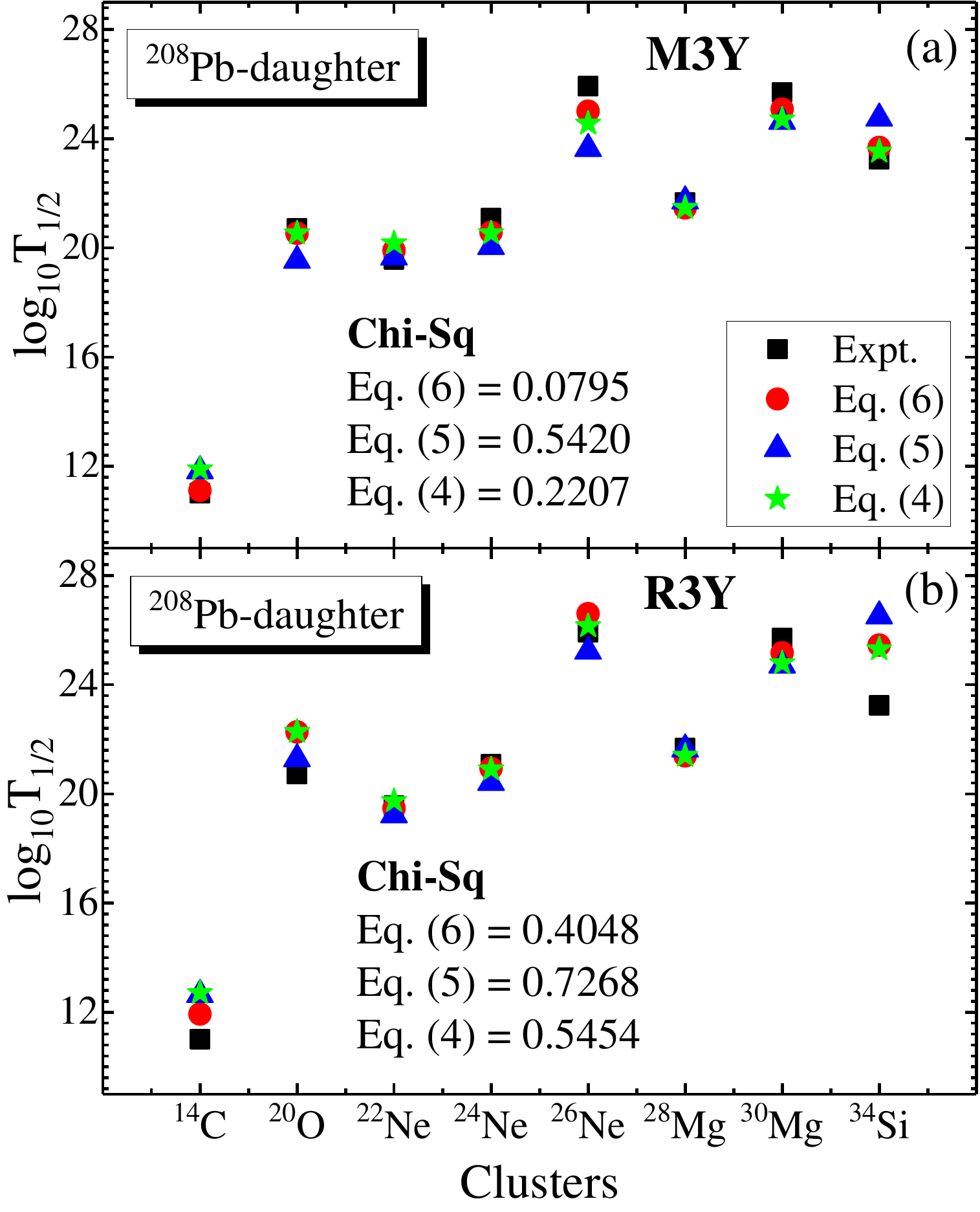}
\caption{\label{fig 1} The logarithmic half-lives of even-even heavy nuclei yielding $^{208}$Pb-daughters calculated using the cluster preformation formula in Eq. (\ref{f1})-(\ref{f3}) and are compared with the experimental data \cite{bone07}.}
\end{figure}

\section{$P_0$ Fitting Procedure}
To compensate for the discussed drawbacks in the $P_0$ formula within the picture of the preformed cluster-decay model, we contemplate three different expressions by considering the inherent features of clusters as the parent nuclei undergo the decay process. From conceptual observations: 
\begin{enumerate}
\item The preformation factor decreases in magnitude with increasing cluster size \cite{sing11}. As such, $P_0$ could be presented in terms of the mass of the cluster $A_c$ and its charge $Z_c$. 
\item The emission of the same cluster from different parent nuclei as well as different clusters from the same parent nucleus is an experimentally observed fact \cite{gupt94,bone07,bone99}. This makes it imperative to consider an expression that shows a dependence on the mass and charge asymmetries $\eta_A=(A_d-A_c)/(A_d+A_c)$ and $\eta_Z=(Z_d-Z_c)/(Z_d+Z_c)$. 
\item The relative distance/separation $r_B$ between the centres of the fragments could play an important role in the formulation of $P_0$. In the touching configuration, it is estimated as  $r_B=1.2(A_c^{1/3}+A_d^{1/3})$ \cite{deli09,qian12}.
\item The Q-value (energy available for the cluster decay process) determines the energetically favourable decay mode and quantitatively influences the preformation quantitatively \cite{isma14}. In addition, the Q value has been shown to reveal information about the nuclear shell structure \cite{deng21}. Moreover, the dependence of the cluster-preformation factor on the Q-value  has been earlier speculated by Qian and  Ren \cite{qian12}.
\end{enumerate}

Based on the known facts in Points: (1-3), we propose the first formula $P_0$ (given the available experimental data). 
\begin{equation}
\log P_0 = -\frac{aA_c\eta_A}{r_B}-Z_c\eta_Z, \label{f1}
\end{equation}
where the constant $a=8.70$ from the nonlinear least square
fitting. The competence of the cluster preformation formula is examined by comparing their respective half-life predictions with the experimental data of the $\chi^2$ function,
\begin{equation}
    \chi^2=\sum^n_{i=1}\frac{\left[\log_{10}T_{1/2}^{cal.}-\log_{10}T_{1/2}^{expt.}\right]^2}{\log_{10}T_{1/2}^{cal.}}.\nonumber
\end{equation}
Using the new $P_0$ formula in Eq.(\ref{f1}), the $\chi^2$  for the considered {\it even-even} systems is estimated to be M3Y =  0.2207  and R3Y = 0.5454 while for 5 odd-A system, the $\chi^2$ is estimated as M3Y = 0.1030 and R3Y = 0.2588 as shown in Table \ref{tab 1}. This expression (Eq. (\ref{f1})) appears promising but does not consider the shell effect which is indispensable to the construction of the preformation probability formula \cite{he21}. Another fact is that a certain amount of energy must be required for the cluster pre-formation process, which is ignored and/or neglected in this expression.  This requires the inclusion of the $4^{th}$ premise, as discussed above.

Since both $\alpha$- and cluster decay emanate from the same physical and theoretical underpinning, we assume that their Q-value dependence similarly follows the Geiger-Nuttall law of $\alpha$-decay \cite{geig11}. Thus, these combine with the mentioned conceptual considerations to yield another $P_0$ formula,
\begin{equation}
    \log P_0 = -\frac{\sqrt{Q}A_c\eta_A}{r_B}-\frac{\sqrt{Z_c}-a}{\eta_Z}-b \label{f2}, 
\end{equation}
and hence the Geiger-Nuttall relationship, $\log_{10}T_{1/2}\propto Q^{-1/2}$ is maintained (since $T_{1/2}=\ln 2/\nu_0P_0P$). Here, $a=5.4808$ and $b=12.6206$ for even-even nuclei. Although the consideration of the Geiger-Nuttall law appears to encompass the known features of nuclear clusters, Eq. (\ref{f2}) may not be the best compromise for estimating the preformation probabilities. This conjecture agrees with the recent investigation of Nagib $\&$ Hamed \cite{nagi21} who demonstrated that the generalization of the Geiger-Nuttall law is rather inadequate for the description of heavy cluster decays. This is probably because the Geiger-Nuttall law remains valid if and only if linearity is maintained between the decay half-life $T_{1/2}$ and the ratio of the square root of the Q-value to the Coulomb barrier height. However, this ratio varies around (0.6-1.0) in cluster emission and thus, non-linearity often becomes inevitable \cite{nagi21}.

\begin{table}
\caption{\label{tab 1} Values of the  nonlinear least-square  fitting parameters $a$, $b$ and $c$ for $P_0$ in Eqs. (\ref{f1}-\ref{f3}) for known experimentally favoured cluster radioactivity (CR) is given in the upper panel. The success of Eq. (\ref{f3}) is extended for the treatment of the theorized heavy particle radioactivity (HPR) in the lower panel. The $\chi^2$ for the half-life predictions for M3Y and R3Y interactions are given in Columns 6 and 7 respectively. Note that odd-odd cluster emitters have not been experimentally observed.}
\vspace{2mm}
\resizebox{0.49\textwidth}{!}{
\centering
\begin{tabular}{|c|c c|c|c|c|c|c|}
\hline
Decay&\multicolumn{2}{c|}{Formula} &\multicolumn{3}{c|}{Constant Parameters}& \multicolumn{2}{c|}{ $\chi^2$} \\
\cline{4-8}
mode& & & a&b&c&M3Y&R3Y\\
\hline
CR&Eq. (\ref{f1}) &e-e &8.70& & &0.2207&0.5454\\
& &o-A &9.89& & & 0.1030& 0.2588\\
& Eq. (\ref{f2}) &e-e&5.48&12.621& &0.5420&0.7168\\
& &o-A & 0.001 &8.504& &0.2484&0.5458\\
& Eq. (\ref{f3}) &e-e& 11.98&0.037&1.52$^\dag$&0.0707&0.1081\\
& & &11.98&0.037&3.56$^\S$&0.0795&0.4048\\
& &o-A &16.12&0.119&4.02&0.0602 &0.1329 \\ 
\hline
HPR$^\P$& &e-e&41.53&0.227&53.93& & \\
 & Eq. (\ref{f3}) &o-e& 65.77&0.269&111.66& &  \\
 & &e-o &75.80&0.194&152.56& & \\
 & &o-o&21.48&0.191&0.001 & & \\
 \hline
\end{tabular}}
\label{tab1}
\footnotesize{\\$^\dag$The value of $c$ is lower for systems with open-shell daughters.}\\
\footnotesize{$^\S$The value of $c$ is higher for systems with closed-shell daughters.}\\
\footnotesize{$^\P$Experimental data is not available.}
\end{table}

Furthermore, the investigation of the cluster emission reveals that the Q-value largely depends on the mass of the participating nuclei (especially, either the parent nucleus or cluster or both). Hence, we contemplate the use of a \textit{weighted  Q-value} such that the emission of the same clusters from different parent nuclei (and vice versa) have their specific quantitative contribution in the $P_0$ expression,
\begin{equation}
     \log P_0 = -\frac{aA_c\eta_A}{r_B}-Z_c\eta_Z+bQ+c, \label{f3}
\end{equation}
where the value of the constant parameters is given in Table \ref{tab 1}. The $\chi^2$ of Eq (\ref{f2}) for the considered {\it even-even} nuclei turn-out to be M3Y = 0.5420 and R3Y = 0.7168 (both higher than those in Eq. (\ref{f1})). The same is true for the deviation of the odd-A nuclei. On the other hand, besides the fact that the expression in Eq. (\ref{f3}) has a relatively lower $\chi^2$ (M3Y = 0.0795 and R3Y = 0.4048) for the same 8 even-even nuclei (and similarly for the 5 odd-A nuclei as shown in Table \ref{tab 1}), it is easy to comprehend and traceable to the physics of cluster emission. Hence, this formula (Eq. (\ref{f3})) has an edge over others (as shown in Fig. \ref{fig 1}). Firstly, it builds a bridge between the decay energy (Q-value) and cluster-preformation probability. Moreover, the third term $bQ$ accounts for the fraction of the decay energy contributed during the cluster preformation process from those required for the complete quantum tunnelling/penetration process. The Q-value is shared among the decaying fragments such that $Q=E_c+E_d$, where the kinetic energy of the emitted cluster is taken as $E_{c}=\frac{A_d}{A}Q$. However, it is assumed that a certain amount of energy must be involved in the formation of clusters before their transmission. Hence, the $3^{rd}$ term in Eq. (\ref{f3}) opens a new paradigm to calculate the actual amount of energy contributed to the cluster formation process from the decay energy
\begin{eqnarray}
Q=\overbrace{\underbrace{ bQ}_{\substack{\text{energy  }\\ \text{contributed in}\\ \text{cluster formation}}}+\underbrace{\kappa\sqrt{Q}}_{\substack{\text{energy }\\ \text{contributed in}\\ \text{cluster emission}}}}^{E_c}+\underbrace{E_d}_{\substack{\text {recoil}\\ \text{energy of}\\ \text{daughter nucleus}}} \label{qexpd}
\end{eqnarray}
where the $\kappa\sqrt{Q}$ is the energy contributed in cluster emission \footnotemark$^,$  \footnotetext{A clue can be taken from the "photoelectric effect" \cite{fan45} in which the energy of photon $\hbar\nu$ = minimum energy needed to remove an electron (work function $\hbar\nu_0$) + kinetic energy of the emitted electron $\frac{1}{2}mv^2$. Thus, $\hbar\nu=\hbar\nu_0+\frac{1}{2}mv^2$. Here, $E_c$ is expressed as the sum of the energy contributed during cluster formation $bQ$ and the energy expended during cluster emission $\kappa\sqrt{Q}$ as described in Eq. \eqref{qexpd}.} \footnotemark\footnotetext{It is also salient to note that the second term $\kappa\sqrt{Q}$  on the right-hand side of  Eq. \eqref{qexpd} mainly contributes to the barrier tunnelling process which plays a vital role in obtaining the decay half-life $T_{1/2}$. This term mimics the correlation between the half-lives of radioactive decay processes and the Q-values in $\alpha$-decay systematics by Geiger and Nuttall  $ \log_{10} T_{1/2}=\kappa \sqrt{Q}-\xi$ \cite{geig11}.}.  
Since,
\begin{eqnarray}
E_c=\frac{A_d}{A}Q=bQ+\kappa\sqrt{Q},
\end{eqnarray}
further simplification gives
\begin{equation}
    \kappa \ = \ \sqrt{Q}\left(\frac{A_d}{A}-b\right). \label{kap}
\end{equation}
The quantity $\kappa$ in Eq.(\ref{kap}) directly influences the penetration probability, which determines the favourable condition for the decay process. It is intriguing to note that the Q-value break-up expression in Eq. (\ref{qexpd}) supports the theoretical description of Greiner \textit{et al.} \cite{mali89,grei86} in which $bQ$ can be correlated with the energy with which cluster formation is initiated at the saddle point within the parent nucleus. Thereafter, it is excited by an internal vibration (which adds to the cluster emission energy $\kappa\sqrt{Q}$) so that the potential at the first turning point is higher than the Q-value. This is immediately accompanied by a de-excitation process following Geiner's ansatz  \cite{grei86} and so, the cluster penetrates across the potential barrier. Secondly, it establishes the interrelationship among the participating parameters, which sheds light on the mechanism involved in the cluster emissions. Furthermore, the formula gives a simple and intuitive path that can be extended to the superheavy region which could be informative for the upcoming experiments on the synthesis of new elements and isotopes. Withal, the properties of nuclei can be classified into two, namely, the universal (global/general) and unique (individualistic) properties \cite{satp87}. For example, all nuclei are composed of nuclear matter, and thus related properties can be reckoned as universal. On the other hand, the unique features of nuclei, namely the shell effect, nuclear halo and/or skin, nuclear shape, shape co-existence, surface diffuseness, and so on, are the exclusive characteristics that differentiate one nucleus from another. The fourth advantage is that the last term (parameter `$c$') of Eq. (\ref{f3}) is connected to shell correlations in terms of open and/or closed-shell nuclei, which is one of the unique properties of the nuclei. This is reminiscent of the parameter $\eta$ of Satpathy {\it et al.} \cite{satp04} and the references therein that accommodate the contributions of various individualistic properties of nuclei. The shell effect can be easily noticed in Table \ref{tab 1} and its footnote as one compares the value of parameter $c$ for even-even nuclei having double magic daughters with its corresponding value for open-shell daughter nuclei. The above prediction has been clarified for given examples of systems producing $^{208}$Pb daughters and non-$^{208}$Pb daughters respectively. The same trend is noticed for other double magic nuclei (which are not shown here for the sake of brevity). Here, other parameters maintained their usual values for the even-even systems. In this respect, this $P_0$ formula encompasses the universal and unique properties of an atomic nucleus.

\begin{figure}
\centering
\includegraphics[scale=0.36]{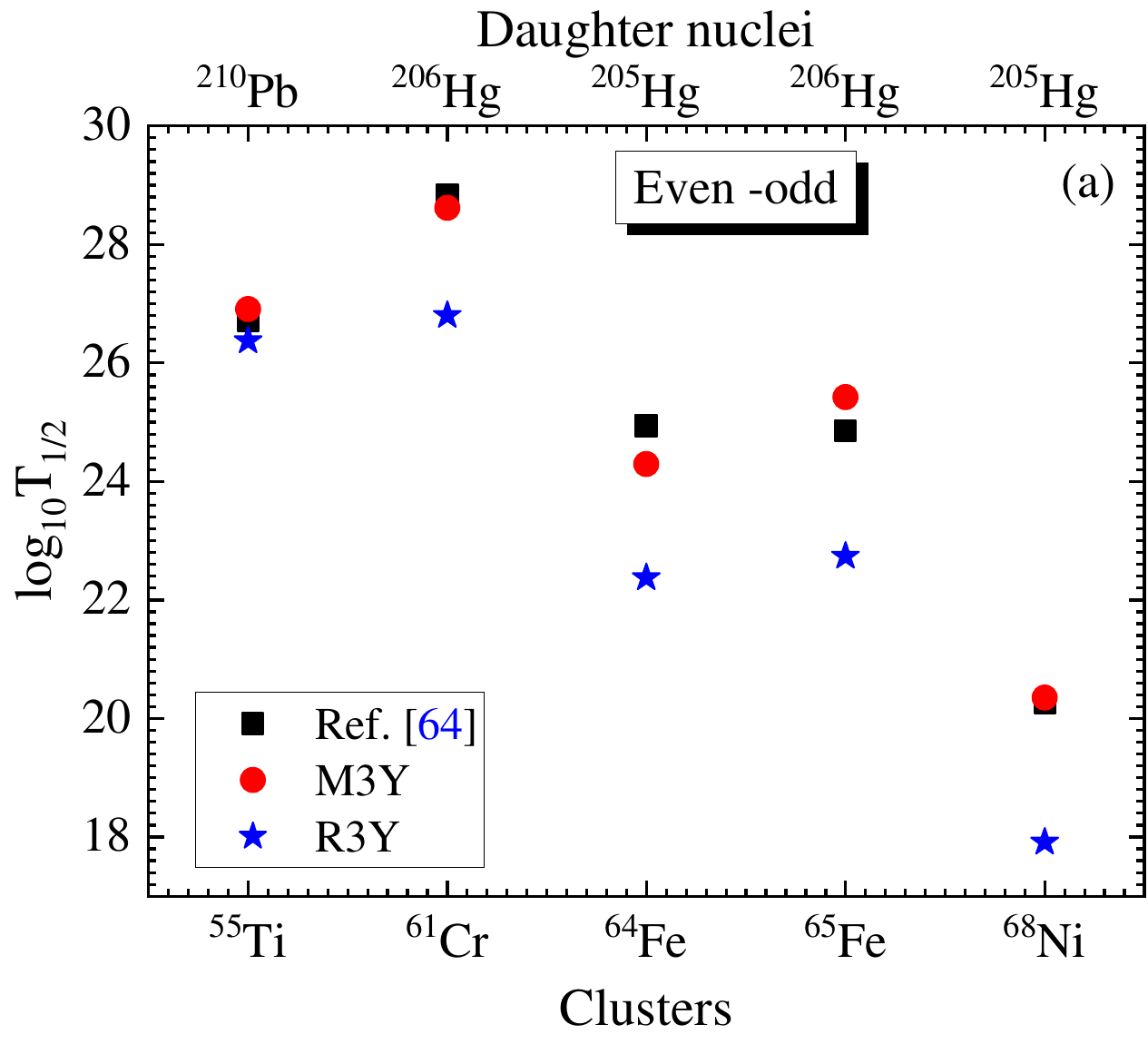} 
\includegraphics[scale=0.36]{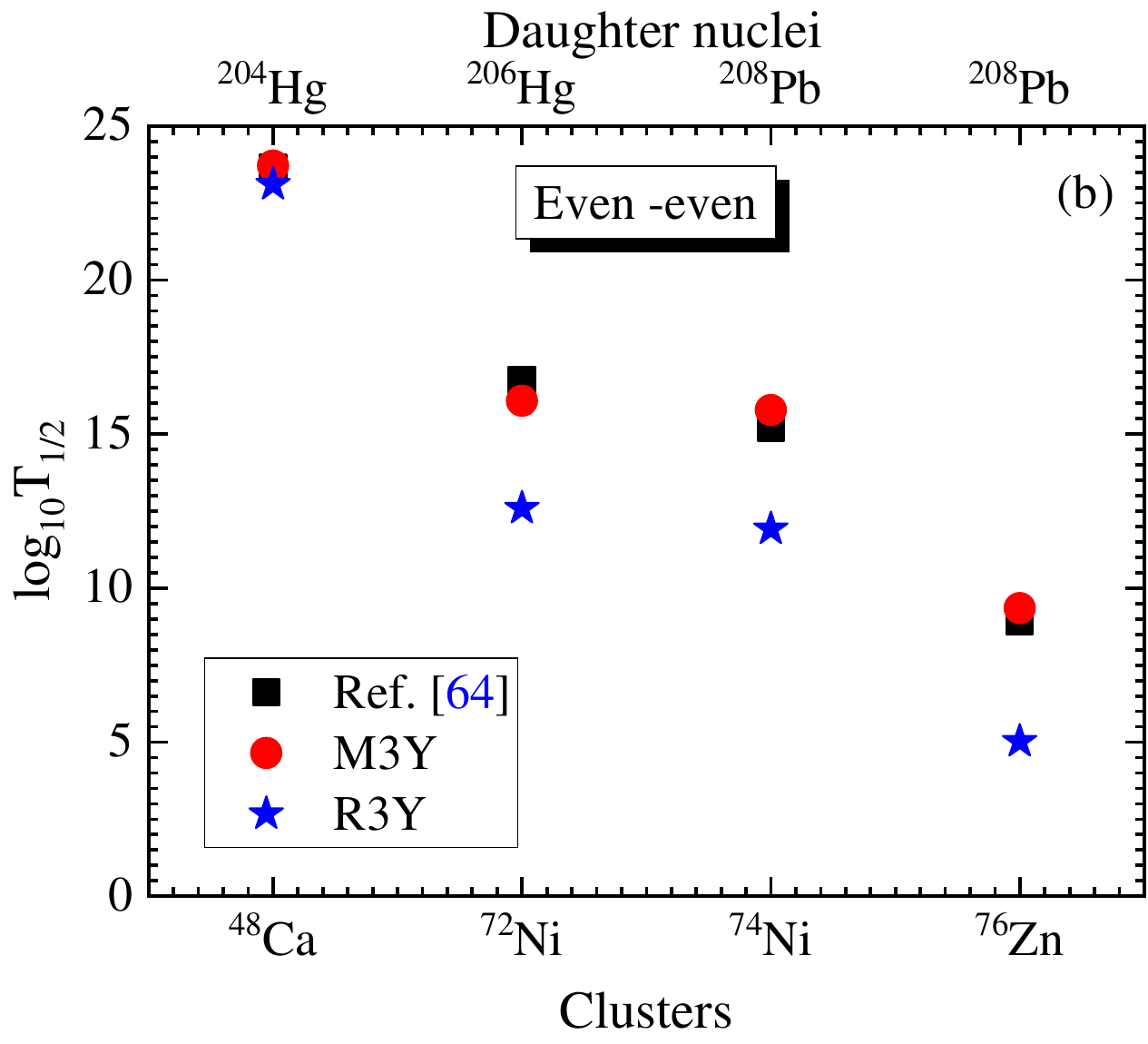}
\caption{\label{fig 2} The logarithmic half-lives of (a) even-odd and (b) even-even heavy clusters calculated using the cluster preformation formula in Eq. (\ref{f3}) for both R3Y and M3Y potentials compared with those of Poenaru {\it et al.} \cite{poen18}.}
\end{figure}
The other two constants $a$ and $b$ in the $P_0$ expression in Eq. (\ref {f3}) may be associated with the inherent features of the associated nuclei, namely, cluster deformation, dynamical effect, quantization, pairing correlation \emph{etc.}, which are yet to be understood. We realized that although the proton $p$ and neutron $n$ of the same nucleus exist in different charge states, the parameters $a$, $b$, and $c$ are associated with approximately equal values for even-odd and odd-even nuclei. This can be ascribed to their similar pairing force despite the arrangement of proton-neutron ($pn$), neutron-proton ($np-$) interactions and perhaps unpaired $n$ or $p$ which has a uniform effect in the estimation of the preformation probability \cite{he21}. On this account, they are represented as odd-A nuclei in Table \ref{tab1}.  Currently, there is no experimental evidence for odd-odd cluster emitters \cite{poen11b}, which are not listed for light and/or intermediate mass cluster radioactivity. The $\chi^2$ estimates for the half-life show that the M3Y and R3Y NN-potential are compatible with the new formula for $P_0$ in Eq. (\ref{f3}). Furthermore, a careful inspection shows that the experimental data and the estimated half-lives from M3Y are a little better as compared to the case of R3Y NN potential. This small disparity can be attributed to the differences in their respective parameterizations. Elaborately, the M3Y NN potential has been phenomenologically fitted to the experimental data of a wide range of nuclei. On the other hand, the R3Y  stems from the microscopic relativistic mean-field Lagrangian where the mesonic degrees of freedom are considered. Nevertheless, both predictions agree well with the experimentally measured half-lives.

Furthermore, it has been demonstrated in Refs. \cite{kuma12c,kuma12s} that the fixed neck-length parameter $\Delta$R = 0.5 fm is most appropriate for cluster decays within the PCM and this is consistently true for the M3Y interaction \cite{jos22b}. Notwithstanding, this is not the case with the recently developed R3Y interaction due to its peculiar barrier characteristics. Hence, for the sake of precision in this fundamental study, the theorized concept of heavy particle radioactivity is approached with the R3Y interaction at the least permissible neck length. Interestingly, using the M3Y, the predictive power of Eq. (\ref{f3}) is found to have higher consistency for HPR in which the emitted fragments are much heavier than the normal clusters emitted during the cluster radioactivity, although they are lighter than the fission fragments. On the other hand, the R3Y predictions are relatively lower than those of M3Y due to its barrier properties, which are more pronounced with increasing cluster size, as earlier demonstrated by Sahu {\it et. al.} \cite{sahu11}. In Fig. \ref{fig 2}, the logarithmic half-lives of various even-even, even-odd superheavy elements are shown corresponding to the emitted cluster are shown. Since the experimental half-lives of HPR are not available at present, hence the comparison is made with the theoretical predictions of Poenaru {\it et al.} \cite{poen18}, which are found to match nicely for M3Y and a relatively large difference for R3Y in terms of $\chi^2$ estimates. However, it is difficult to conclude the compatibility and/or validity of the models due to lack of the experimental data for HPR.

\section{Summary and  Conclusions}
In conclusion, the newly proposed formula captures the influential parameters of the cluster radioactivity and its predictive power spans the entire nuclear territory. This formulated expression is not vaguely proposed by mere fitting to some arbitrary values but rather based on established conceptual findings in the cluster emission whose interrelations are studied such that it describes and accommodates the physics of cluster preformation and its mechanisms. For the first time, the Q-value is modelled to give a quantitative description of the energy contributed during the cluster preformation process and the recoil effect of the daughter is appraised, ensuring that the energy is conserved. Interestingly, the formula adapts reasonably well to both microscopic and phenomenological potentials, and all calculated half-lives agree well with the existing theoretical models and available experimental data. Thus, we have demonstrated that this formula supports the concept of heavy-particle radioactivity and could throw new light on the possibility of clustering in superheavy nuclei, which is a current topic of interest in the nuclear physics.

\acknowledgments
This work is supported by the Ministry of Education Malaysia, Grant No: FRGS/1/2019/STG02/UNIMAP/02/2), Science Engineering Research Board (SERB), File No. CRG/2021/001229, FOSTECT Project Code: FOSTECT.2019B.04, and FAPESP Project Nos. 2017/05660-0.

\end{document}